\documentclass[aps,pra,twocolumn,10pt,a4paper,floatfix,footinbib]{revtex4}
\usepackage{amssymb}
\usepackage{graphicx}
\usepackage{epsfig}
\usepackage{amsmath}
\usepackage{slashed}
\usepackage{units}
\usepackage{hyperref}
\usepackage{array}
\usepackage{lmodern}
\usepackage{dcolumn}
\usepackage{bm}
\usepackage{tabularx}
\usepackage[latin1]{inputenc}

\immediate\write18{bibtex \jobname}

\begin{document}

\title{Quantum vacuum photon modes and repulsive Lifshitz-van der Waals interactions}

\author{Louis Dellieu$^1$}
\email{louis.dellieu@unamur.be}
\affiliation{Research Center in Physics of Matter and Radiation (PMR), Department of Physics, University of Namur, 61 rue de Bruxelles, B-5000 Namur, 
Belgium}

\author{Olivier Deparis}
\affiliation{Research Center in Physics of Matter and Radiation (PMR), Department of Physics, University of Namur, 61 rue de Bruxelles, B-5000 Namur, 
Belgium}

\author{J\'{e}r\^{o}me Muller}
\affiliation{Research Center in Physics of Matter and Radiation (PMR), Department of Physics, University of Namur, 61 rue de Bruxelles, B-5000 Namur, 
Belgium}

\author{Branko Kolaric}
\affiliation{Research Center in Physics of Matter and Radiation (PMR), Department of Physics, University of Namur, 61 rue de Bruxelles, B-5000 Namur, 
Belgium}

\author{Micha\"{e}l Sarrazin$^1$}
\email{michael.sarrazin@unamur.be}
\affiliation{Research Center in Physics of Matter and Radiation (PMR), Department of Physics, University of Namur, 61 rue de Bruxelles, B-5000 Namur, 
Belgium}

\begin{abstract}
The bridge between quantum vacuum photon modes and properties of patterned surfaces is currently being established on solid theoretical grounds. Based on these foundations, the manipulation of quantum vacuum photon modes in a nanostructured cavity is theoretically shown to be able to turn the Lifshitz-van der Waals forces from attractive to repulsive regime. Since this concept relies on surface nanopatterning instead of chemical composition changes, it drastically relaxes the usual conditions for achieving repulsive Lifshitz-van der Waals forces. As a case study, the potential interaction energy between a nanopatterned polyethylene slab and a flat polyethylene slab with water as intervening medium is calculated. Extremely small corrugations heights (less than ten nanometers) are shown to be able to turn the Lifshitz-van der Waals force from attractive to repulsive, the interaction strength being controlled by the corrugation height. This new approach could lead to various applications in surface science.\\
$^1$\textit{These authors have contributed equally to this work.}
\end{abstract}

\maketitle
\section{Introduction}

Many fundamental aspects and practical issues in the physics of interfaces are related to controlling interactions between surfaces \cite{1,2}. In his seminal article \cite{3}, De Gennes pointed out the importance of van der Waals and electrostatic forces in adsorption, adhesion and wetting phenomena. Recently, it became obvious that controlling forces between macroscopic bodies or surfaces is crucial for a variety of applications such as mechanics of nanomachines, stability of colloids and communication between biological cells, for instance \cite{1,2,4,5}.

The growing interest for nanoelectromechanical systems (NEMS) urges the scientific community to deeply study van der Waals and electrostatic interactions within nanostructured systems \cite{5a,5aa}. In particular, looking at nanostructures in the theoretical framework of dispersive (van der Waals) interactions turns out to be of great interest, from both fundamental perspectives and quantum based-technologies \cite{5aaaaa}. In order to understand the influence of surface corrugations on Lifshitz-van der Waals interactions between macroscopic bodies, many approaches have emerged, each one addressing specific corrugation geometries \cite{5aa}. On one hand, additive methods, such as the proximity force approximation (PFA) or the pairwise summation (PWS), appear to be the most employed ones while describing interactions between smooth corrugated surfaces at short and long separation distances, respectively \cite{5aa,5prim}. On the other hand, while considering corrugated surfaces with small correlation length (of the order of the separation distance), nonadditive methods such as the scattering or perturbative approaches are required in order to take into account diffraction and correlation effects which occur at the nanoscale \cite{5aa,5prim2}. However, it is noteworthy that refinements of the above mentioned approaches have led to their progressive convergence for specific corrugations size and shape \cite{5aa}.

This article focuses on small correlation length nanostructures with steep features. In this case, the assimilation of the surface corrugation to a graded effective refractive index layer is relevant \cite{5aaa, 5aaaa, 17} and allows simplifying greatly the application of nonadditive methods, while still taking fully into account electrodynamical coupling between features as it will be explained below. Motivated by recent theoretical \cite{5bb,20a,17} and experimental \cite{5b} studies, we introduce the novel concept of manipulating quantum vacuum photon modes at the sub-10nm scale in order to turn Lifshitz-van der Waals interactions from attractive to repulsive.

The article is organized as follows. After introduction (Section I), we present in Section II the theoretical framework used for the description of the Lifshitz-van der Waals interactions and we explain our theoretical approach for calculating the interfacial energy in the case of steep nanocorrugated surfaces. The relevant approximation used to describe
steep nanocorrugated surfaces is introduced and justified in the Section III. In Section IV we report on computational results and discuss how sub-10 nm corrugations allows us to control quantum vacuum photon modes and, by this way, to turn the Lifshitz-van der Waals force from attractive to repulsive. Perspectives and general remarks are finally provided in Section V.

\begin{figure}[b]
\centerline{\ \includegraphics[width=8 cm]{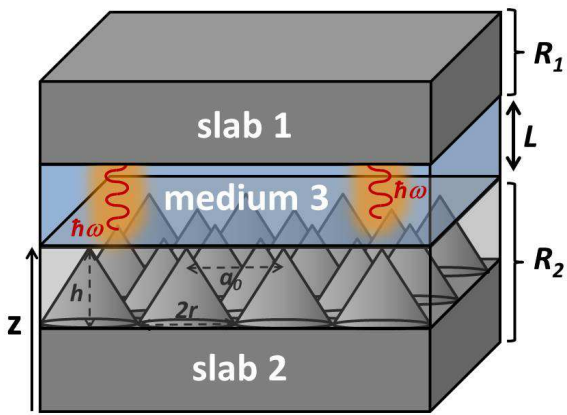}}
\caption{(Color online). Bodies $1$ and $2$ interacting \textit{via} an intervening medium $3$. The two bodies are separated by a distance \textit{L}. $R_1$ ($R_2$) is the Fresnel reflection coefficient of slab $1$ (slab $2$). The surface of body $2$ is nanopatterned with corrugations that are described by a graded effective medium.}
\label{fig1}
\end{figure}

\section{Theoretical approach}
Over the last decades, the Lifshitz - van der Waals approach of
interfacial interactions in macroscopic systems has been widely investigated, both
theoretically and experimentally \cite{6,7,7b,8,8b,8c,8d}. Although usually attractive, the
interaction potential energy may become repulsive if particular conditions
are satisfied \cite{9}. Let us first consider a body (1) interacting with a body (2) \textit{via} an
intervening medium (3) (Fig. 1).
Let us also consider for the moment that each element of this macroscopic system has a planar geometry (for now body (2) is treated as a
flat homogeneous slab) and is characterized by a dielectric function $\varepsilon $(i$\xi $),
where i$\xi $ is the imaginary angular frequency. It is well established that
repulsive Lifshitz-van der Waals interactions between the two bodies can take place if the following condition
is satisfied \cite{9}:
\begin{equation}
\varepsilon _1(i\xi ) < \varepsilon _3(i\xi ) < \varepsilon _2(i\xi )
\end{equation}
where $\varepsilon _i$ is the dielectric function of the i$^{th}$ component
of the system. 
Eq. (1) cannot be satisfied if the intervening medium is vacuum. Therefore, a liquid or a gas is needed to satisfy Eq. (1) 
for given slab materials \cite{9}. Moreover, in practice, Eq. (1) imposes tight 
constraints on the choice of both the materials 
and the intervening medium, which makes challenging the experimental observation of repulsive Lifshitz-van der Waals interactions \cite{10,11}.

Hereafter, we introduce an original approach to modify Lifshitz-van der Waals
interactions in a very different perspective, beyond the constraint set by Eq. (1). By tuning the virtual photon exchange
between the two bodies, \textit{via} nanopatterning of the surface of one of them (Fig. 1), it is possible
to obtain a repulsive interaction potential energy, without any
modification of the chemistry of materials, \textit{i.e.} without changing their dielectric functions. The nature of the Lifshitz-van der Waals force -- repulsive or
attractive -- is solely the result of controlling
light-matter quantum interactions at the nanoscale. 

The concept briefly described above is based on the fundamental interplay between
physics of confined media and optical cavities. 
Here, the concept of confined space is applied to the particular geometry of a
planar-like cavity \cite{12,13,14} (Fig. 1).
Actually, the present approach relies on joined effects
of electromagnetic confinement and surface patterning at the nanoscale which are exploited to
modify the interaction potential energy. In addition, since the studied system can be
regarded as an optical cavity, it is possible to establish a formal link between optical
properties of the cavity (quality factor $Q$ for instance) and Lifshitz-van der Waals forces between the slabs forming that cavity.

At the macroscopic level, the force, thus the interaction potential energy,
between two planar surfaces is related to the lowest (zero-point) energy state of the
electromagnetic (EM) field, arising from the existence of virtual photons of energy 
$\frac 12\hbar \omega $ at all available frequencies which are exchanged between both surfaces \cite{5aa}. The interaction
potential energy \textit{U} can be written as \cite{16b} $U(L)=\frac 12\sum_k\hbar (\omega
_k(L)-\omega _k(L\rightarrow \infty ))$, where $\omega _k$ is the angular
frequency of the k$^{th}$ vacuum photon-mode available between the two surfaces
separated by a length \textit{L} (Fig. 1). $U(L)$ can be then easily related to the density of
EM states $\rho (\omega ,L)$ of the system such that: $U(L)=\frac 12\hbar \int
\omega (\rho (\omega ,L)-\rho (\omega ,L\rightarrow \infty ))d\omega $.
The quantity $\rho (\omega ,L)$ can be obtained from classical electrodynamics. 

The tuning of the zero-point energy is possible thanks to the presence of surfaces
(boundaries), \textit{i.e.}, the presence of allowed modes of the EM field within the cavity \cite{5,5aa}. Taking this fact into
account, the force appears at the macroscopic level as the result of a manifold of vacuum
photon modes occurring because the
EM field must meet the appropriate boundary conditions at each
surface. Moreover, these vacuum photon modes can be altered by patterning the surfaces \cite{12}, i.e. EM field boundaries.

As explained above, the interaction potential energy considered in the present case results from
the exchange of virtual photons between two interacting bodies. By summing the individual energies related to each mode available
within the cavity, we can retrieve the total
energy of the system from the photon density of states. Based on these arguments, we apply the so-called scattering approach \cite{5aa} to calculate the interaction potential energy
between two bodies facing each other. Accordingly, the interaction potential energy \textit{U} is given by:

\begin{eqnarray}
U(L) &=&\frac \hbar {2\pi }\sum_{m=s,p}\int \frac{d^2k_{//}}{(2\pi )^2}%
\int_0^\infty d\xi  \label{3} \\
&&\times \ln (1-R_1^m(i\xi ,\mathbf{k}_{//})R_2^m(i\xi ,\mathbf{k}%
_{//})e^{-2\kappa L})  \nonumber
\end{eqnarray}
where \textit{L} is the separation distance between the bodies, $\kappa =\sqrt{\frac{%
\xi ^2}{c^2}+|k_{//}|^2}$ , $R_1^m(R_2^m)$ is the generalized complex
reflection coefficient of the first body (second) in the \textit{m} polarization
state (\textit{s} or \textit{p} states), $k_{//}$ is the parallel component of the photon wave
vector and $i\xi $  is the imaginary angular frequency \cite{rosa}. It is noteworthy that, in this method,
the nanopatterned slab (here body (2)) is treated as a graded effective medium (see \cite{17} and discussion below).

For short separation distances (\textit{L} $\leq $ 10 nm), Eq. (2) is well approximated by the
so-called Hamaker formula \cite{18}:
\begin{equation}
U(L)=-\frac{A_{132}}{12\pi L^2}
\end{equation}
where $A_{132}$ is the effective Hamaker constant of the system which can be
deduced from the numerically computed energy, \textit{i.e.}, Eq. (2).

In case the intervening medium is vacuum, Eq. (2) appears to be efficient since it reproduces well
experimental results \cite{17}. In this case, using Eqs. (2) and (3), it is possible to retrieve
the Hamaker constant $A_{12}$ of the system ($A_{12}$ $\equiv A_{132}$, where (3)
is omitted when the intervening medium is vacuum). Using the same procedure, it is also possible to retrieve the Hamaker
constant $A_{11}$ of the flat surface. The
effective Hamaker constant $A_{22}$ of the patterned surface, on the other hand, can be deduced indirectly
from the well-known relation \cite{2} $A_{12}=\sqrt{A_{11}}\sqrt{A_{22}}$ by
using values of $A_{12}$ and $A_{11}$ calculated from Eqs. (2) and (3). 

For numerical convenience \cite{visser,visser2}, instead of computing directly $A_{132}$ via Eq. (2) 
to obtain the interaction potential energy, we compute the effective Hamaker constant of the
system (hereafter, (3) stands for fluid) from the well-known relation \cite{2,visser,visser2}:

\begin{equation}
A_{132}=(\sqrt{A_{11}}-\sqrt{A_{33}})(\sqrt{A_{22}}-\sqrt{A_{33}})
\end{equation}
where $A_{ii}$ are the Hamaker constants of the corresponding media
which are obtained from the above described procedure.
Repulsive interaction potential energy is reached when $A_{132}$ is negative, \textit{cf}. Eq. (3).
Therefore, according to Eq. (4), such a condition is fulfilled when :

\begin{equation}
A_{22} < A_{33} < A_{11} .
\end {equation}

The condition imposed by Eq. (5) goes beyond the constraint set by Eq. (1). Indeed, when considering a nanopatterned surface, as in the present case, Eq. (1) can not be used
since dielectric functions are those of flat materials. On the other hand, Eq. (5) allows to bypass this problem since the effective Hamaker constant of the nanopatterned
surface can be calculated by the above described procedure. Therefore, Eq. (5) has a more general application since it can be used simultaneously for both flat and nanopatterned surfaces.

\begin{figure}
\centerline{\ \includegraphics[width=8.5 cm]{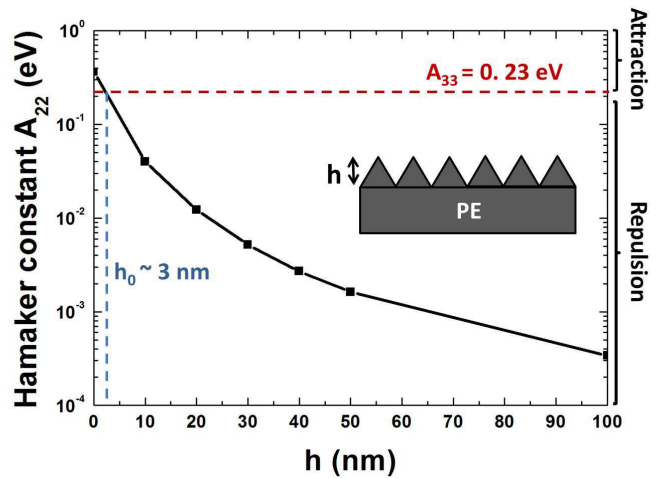}}
\caption{(Color online). Hamaker constant $A_{22}$ of nanopatterned polyethylene slab (see inset) as a function of cones height $h$. }
\label{fig2}
\end{figure}

\section{Modelling and simulation details}
Let us now develop the case of a practical two-body system consisting of two polyethylene (PE) slabs facing each other and separated by
a \textit{L} distance. In order to fulfill Eq. (5), we choose water (medium 3) as intervening medium. 
It must be pointed out that such a configuration does not match Eq. (1).
The first slab (slab 1) has a flat (planar)
surface while the second one (slab 2) is nanostructured with cones of height \textit{h}
arranged on a hexagonal lattice with a lattice period chosen to
be $a_0$ = 10 nm (Fig. 1). We choose a fixed cone base radius of $r$ = 5 nm and
a variable cone height \textit{h} (ranging from 10 to 100 nm) in order to alter the
optical properties of the surface. Indeed, such a geometry is
known to improve the antireflection behavior of the surface which in turns alters
the vacuum photon modes of the system \cite{17,19}. Moreover, since the PE surface is hydrophobic \cite{1},
we can assume a Cassie state \cite{19b} between water and the corrugated PE surface \cite{17}. As a consequence,
the void space between cones is filled by air and water is localized above the top of the cones only. 

At wavelengths below 20 nm, PE permittivity is close
to unity \cite{20} and thus only vacuum modes above this spectral range are relevant. Since the lattice period is shorter than the relevant wavelength range, the patterned surface can be described by an
effective material, \textit{i.e.} an effective medium approach (EMA), with a graded permittivity $\varepsilon _{eff}(z)$ 
along its thickness such that: 
\begin{equation}
\varepsilon _{eff}(z)=1+(\varepsilon _{PE}-1)f(z)
\end{equation}
where $\varepsilon _{PE}$ is the PE dielectric function,  $f(z)$ is the
filling fraction given by $f(z)=\pi r(z)^2/S$ with $S=a_0^2\sqrt{3}/2$ 
and $r(z)$ the radius of the circular section of the cones at coordinate
\textit{z}.

The use of the EMA has to be justified with great care. Indeed, such an approximation usually requires the separation distance \textit{L} to be equal or larger than the lattice parameter $a_0$ of the periodically nanostructured surface \cite{20a}. The reason is that the distance is one of the main parameter determining the nature of the electromagnetic modes (radiative or evanescent) involved in the calculation of the Lifshitz-van der Waals interaction. In the case of a separation distance shorter than the lattice parameter, evanescent modes are dominant and are able to reproduce the details of the nanostructure, thereby invalidating the EMA \cite{20b,20c}. However, in the present situation, the EMA remains valid for the separation distances shorter than the lattice parameter. This non-intuitive result emerges from the weakness of the coupling between diffracted and specular orders due to both the optical properties of PE and the steepness of the corrugation
(see Appendix A for a detailed justification of the use of EMA).

The impact of the choice of water as the intervening medium has to be examined. Indeed, as a polar liquid, water induces an electrostatic double layer at both surfaces facing each other, giving rise to an additional electrostatic repulsive force between them which, in experiments, could screen the Lifshitz-van der Waals repulsive force treated here. However, while considering the Cassie state regime, the electrostatic double layer is located only on the top of the cones which become steeper while increasing the cones height. As a consequence, the electrostatic double layer associated to the steep nanocorrugated PE surface becomes extremely small, leading to a dramatic decrease of the electrostatic repulsive interaction (see Appendix B). Therefore, in the present situation, the electrostatic repulsive interaction can be neglected compared to the Lifshitz-van der Waals interaction for any corrugations height.

The corner stone of the present approach is related to the fact that tuning
antireflective properties of the bottom slab (thanks to nanocorrugations) allows tailoring the virtual photon exchange within
the cavity formed by the top flat PE slab and the bottom corrugated PE
slab \cite{17}. Since the presence of virtual photons causes dispersive interaction energy between slabs, it enables the control of the magnitude of
attraction/repulsion between the two surfaces by only playing on the photon mode
density inside the cavity.

\begin{figure}
\centerline{\ \includegraphics[width=8.5 cm]{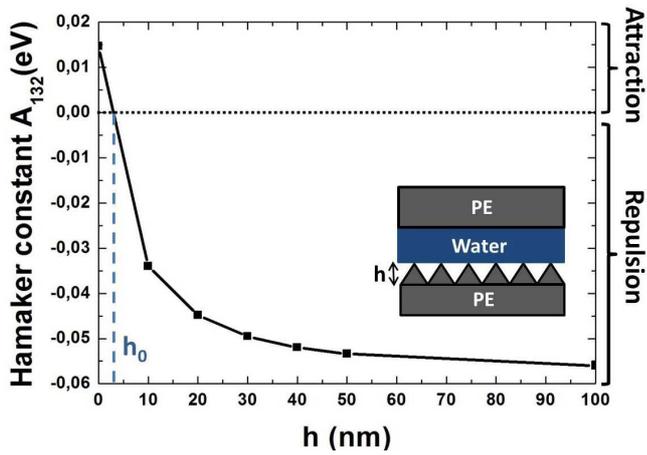}}
\caption{(Color online). Hamaker constant $A_{132}$ of flat polyethylene/patterned polyethylene system immersed in water (schematic view in the inset) as a function of cones height $h$.}
\label{fig3}
\end{figure}

\section{Results and discussion}
As explained previously, the use of a fluid as intervening medium led no choice but to compute individual Hamaker constants of each component of the system
in a first step. We calculate from previously reported data \cite{17} the
effective Hamaker constant $A_{22}$ of the corrugated PE slab using Eq. (2)
and Eq. (3) for various cones height (Fig. 2). Knowing the
Hamaker constant $A_{11}$ of flat PE surface ($A_{11} = 0.36$ eV) \cite{22}  and
the Hamaker constant $A_{33}$ of water ($A_{33} = 0.23$ eV) \cite{2}, we then
calculated the Hamaker constant $A_{132}$ of the whole system
from Eq. (4), as a function of the cones height (Fig. 3).
Strong decrease of the Hamaker constant $A_{132}$
with increasing cones height is observed, going from positive to negative
values (Fig. 3). Here, the zero-crossing point for $A_{132}$ takes place for $h_0 \approx 3$ nm.
This critical point is reached when $A_{22} = A_{33}$ (horizontal red-dotted line, Fig. 2),
in accordance with the fact that repulsive
interaction is achieved only if Eq. (5) is satisfied.
Therefore, the interaction potential energy
becomes positive, \textit{i.e.} repulsive force (Fig. 4), for $h > h_0$ as soon as the Hamaker constant $A_{132}$
of the system becomes negative (see footnote \cite{22bb}). It is noteworthy that, due to the small zero-crossing point value $h_0$, 
it could be experimentally difficult to achieve a progressive transition from attractive to repulsive force while increasing the cones height (see footnote \cite{22b}, \cite{22a}).
Thus, observation of this progressive transition would require higher $h_0$ value
which could be achieved by using flat materials with Hamaker constants $A_{22}$ ($h$ =0) higher than the Hamaker constant $A_{33}$ of the intervening medium (by at
least one order of magnitude, see Fig. 2). Furthermore, owning to the Cassie state regime of the present model, such materials are difficult to find \cite{1}.
Consequently, the transition from attractive to repulsive regimes cannot be easily observed as well as significantly modified (\textit{i.e.}
$h_0$ value stays in the same range for most of the materials).

\begin{figure}
\centerline{\ \includegraphics[width=8.5 cm]{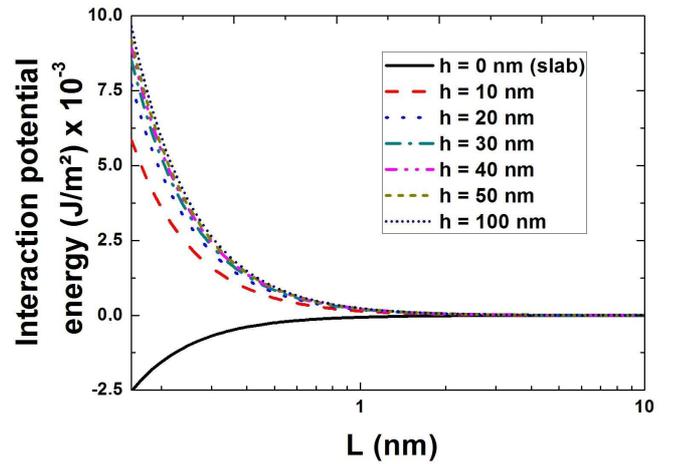}}
\caption{(Color online). Interaction potential energy between PE slabs as a function of cones height.}
\label{fig4}
\end{figure}

Such a dramatic modification of the dispersive energy of the system arises from the
strong decrease of the Hamaker constant $A_{22}$ of the nanopatterned PE slab as the cone height increases (note the logarithmic scale in Fig. 2). Phenomenologically,
this result can be explained by the fact that the increase of the cone
height $h$ causes the decrease of the reflection coefficient of the
nanopatterned PE slab $\mathcal{R}_2$. Consequently, the quality factor 
\textit{Q} of the Fabry-Perot cavity also
decreases (Fig. 5a) since \cite{23}:

\begin{equation}
Q=-2\pi \frac 1{\ln (\mathcal{R}_1\mathcal{R}_2(1-\mathcal{A}_3)^2)}\frac{2L}\lambda 
\end{equation}
where $\mathcal{R}_i = |R_i|^2$ and $\mathcal{A}_3$ is the optical absorption loss of the intervening medium (3) for a single path in the cavity (i.e. $\mathcal{A}_3$ is given by Beer-Lambert law).
\begin{figure}
\centerline{\ \includegraphics[width=8.5 cm]{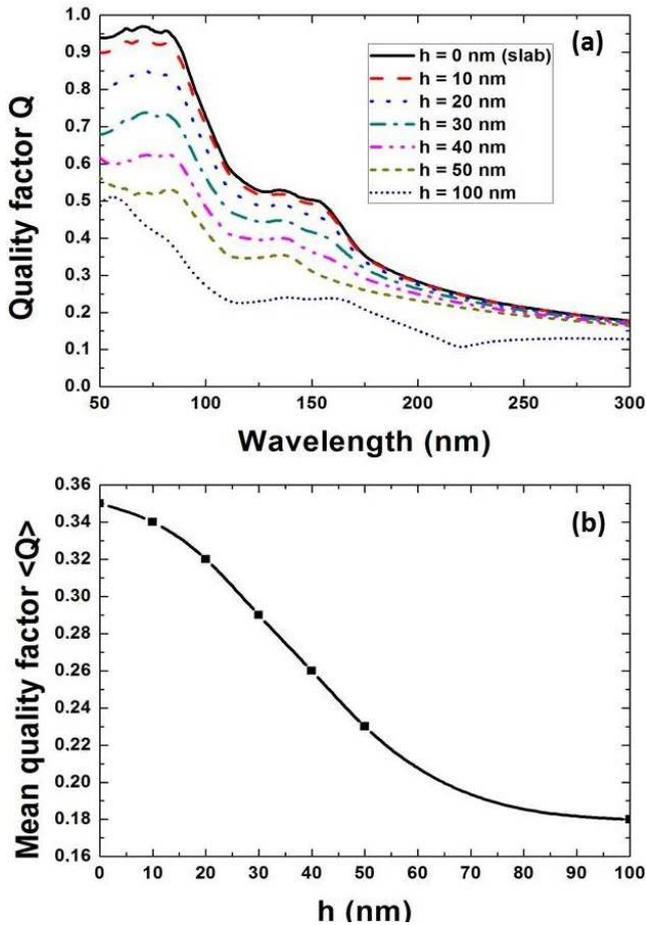}}
\caption{(Color online). (a) Quality factor Q of the Fabry-Perot cavity formed by the flat polyethylene slab and the patterned polyethylene slab. \textit{Q} is given against the wavelength for various cone heigths. (b) Mean value (integration over wavelength) of the quality factor of the cavity described above. Both slabs are separated by water.}
\label{fig5}
\end{figure}
Moreover, the mean value of the quality factor $<Q>$ was calculated by integration of Eq. (7) in the relevant spectral range from 50 nm to 300 nm (Fig. 5b). We observe a decrease of $<Q>$ as the cones height increases, down to 49 \% of the initial value (\textit{i.e.} for flat slabs) (Fig. 5b). As \textit{Q} decreases, the EM
energy stored into the Fabry-Perot cavity is reduced \cite{23}; \textit{i.e.}, the number of vacuum
photon modes available within the cavity and which contribute to the interaction potential energy $U(L) \propto \sum_k\hbar \omega
_k(L)$ diminishes. Therefore, the attractive behavior of the interaction
potential energy becomes weaker while, owing to the geometry of the studied system and the choice of materials, the repulsive behavior becomes stronger. On overall, controlling the optical properties
of the cavity enables tuning the strength of the attractive/repulsive force. It is noteworthy that such an interpretation, although based on solid physical ground (i.e. the energy storage in an optical cavity), relies on a heuristic approach and still requires the establishment of a direct theoretical link between the quality factor and the Lifshitz-van der Waals interaction.

\section{Conclusion}
In conclusion, we showed theoretically that extremely small and steep nanoscopic
corrugations on the surface of one of the two interacting bodies are able to turn Lifshitz-van der Waals interaction from attractive to
repulsive, as well as to control
the strength of their interaction by changing the corrugations height. The present approach is appealing
since it offers the possibility to achieve repulsive interaction only by nanopatterning
one of the surfaces. Therefore, constraints with respect to the choice of materials \cite{9,10,11} are relaxed. 
In the end, we are aware of the fact that an experimental proof of the concept is a very difficult task due to engineering complexity related to the achievement of the pattern depth of just a few nanometers. However, in the light of a recent experimental study \cite{5b}, the presented concept could open new perspectives to control attractive/repulsive interactions by nanopatterning.
Indeed, fabricated patterns could
be used to control macroscopic interactions in a variety of applications
ranging from biology \cite{25} to material science for
controlling wettability, adhesion and adsorption \cite{1,2}.

\textbf{Acknowledgments}

L.D. is supported by the Belgian Fund for Industrial and Agricultural
Research (FRIA). M.S. is supported by the Cleanoptic project (development of
super-hydrophobic anti-reflective coatings for solar glass panels/convention
No.1117317) of the Greenomat program of the Wallonia Region (Belgium). B. K. acknowledges financial support from the Action de Recherche Concert\'ee (BIOSTRUCT project) of the University of Namur (UNamur) and the support from the Nanoscale Quantum 
Optics COST-MP1403 action.
This
research used resources of the ``Plateforme Technologique de Calcul
Intensif'' (PTCI) (http://www.ptci.unamur.be) located at the University of
Namur, Belgium, which is supported by the F.R.S.-FNRS. The PTCI is member of
the ``Consortium des Equipements de Calcul Intensif (CECI)''
(http://www.ceci-hpc.be).

\appendix

\section{Justification of the effective medium approach}

Although the effective medium approach (EMA) is fully relevant in optics while
dealing with the far field, it could be objected that it is inappropriate to
describe the near field, which plays a non trivial role in the calculation of the interaction potential energy at the
short distances\cite{nearfield,plasmon}. Indeed, the evanescent near field exhibits lateral fluctuations which are expected to mimic the
surface corrugation \cite{nearfield2}.
Therefore, EMA is \textit{a priori} inappropriate in describing accurately these fluctuations.
However, the careful analysis described below leads to qualify this restriction in the specific case of subwavelength periodically patterned surfaces.
\begin{figure}[b]
\centerline{\ \includegraphics[width=7cm]{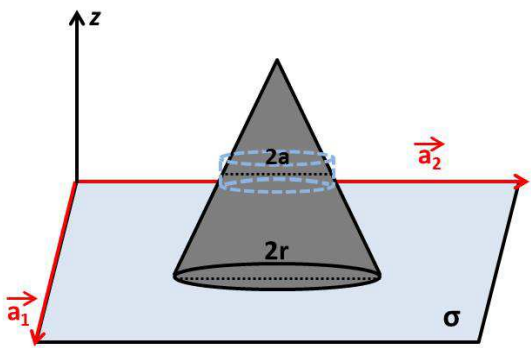}}
\caption{(Color online). Primitive cell of the periodic patterned structure under study. In computations, the cone is described by a stack of cylinder of radius $a$ $\in$ [0,$r$].}
\label{fig2}
\end{figure}
The specular order couples with all the evanescent diffracted orders, which are all coupled together as well. As a result, they constitute the fluctuating near field.
Let us formally examine such couplings in the theoretical framework of the Rigorous Coupled Wave Analysis (RCWA) method \cite{rcwa}. The Fourier series expansion of the dielectric constant writes as :
\begin{equation}
\varepsilon (z,\mathbf{\rho })=\sum_{\mathbf{g}}\varepsilon
_{\mathbf{g}}(z)e^{i\mathbf{g}.\mathbf{\rho }}
\end{equation}
where $\mathbf{\rho \text{ }}$ denotes a real-space vector in the primitive cell 
with basis vectors $%
\mathbf{a_1}$ and  $\mathbf{a_2}$ (Fig. 6), and $\mathbf{g}$ is a reciprocal lattice  vector.
\begin{figure}[b]
\centerline{\ \includegraphics[width=8 cm]{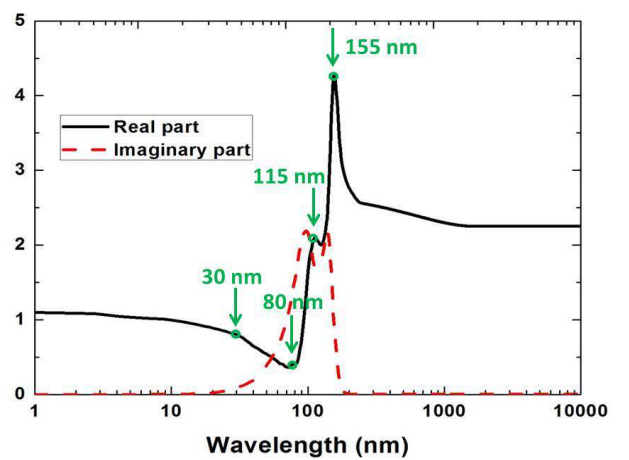}}
\caption{(Color online). Dielectric function $\varepsilon_s$ of PE. Pointed values are those used for numerical simulations (see Fig. 9).}
\label{fig2}
\end{figure}
By virtue of Floquet-Bloch theorem,
the electric $\textbf{E}$ and displacement $\textbf{D}$ fields expand as :

\begin{equation}
\mathbf{E}(z,\mathbf{\rho }) =\sum_{\mathbf{g}}
\mathbf{E}_{\mathbf{g}}(z)e^{i(\mathbf{g}+
\mathbf{k}_{//}).\mathbf{\rho }}
\end{equation}

\begin{equation}
\mathbf{D}(z,\mathbf{\rho }) =\sum_{\mathbf{g}}
\mathbf{D}_{\mathbf{g}}(z)e^{i(\mathbf{g}+
\mathbf{k}_{//}).\mathbf{\rho }}
\end{equation}
Since $\mathbf{D}=\varepsilon _0\varepsilon (z,\mathbf{\rho }%
)\mathbf{E}$, we can re-write $\mathbf{D}_{\mathbf{g}%
}(z)$ as :
\begin{equation}
\mathbf{D}_{\mathbf{g}}(z)=\sum_{\mathbf{g^{\prime }}%
}\varepsilon _0\varepsilon _{\mathbf{g},\mathbf{g^{\prime }}}%
\mathbf{E}_{\mathbf{g^{\prime }}}(z)
\end{equation}
where $\mathbf{g^{\prime }}$ is another reciprocal
lattice vector  and the Fourier matrix element  $\varepsilon _{\mathbf{g},\mathbf{g^{\prime }}}$
expresses the coupling between diffracting orders $%
\mathbf{g}$ and  $\mathbf{g^{\prime }}$. When describing the cones by cylinders stack,  $\varepsilon _{\mathbf{g},\mathbf{g^{\prime }}}$ in a given layer 
is written as :
\begin{equation}
\varepsilon _{\mathbf{g},\mathbf{g^{\prime }}}=\varepsilon
_m\delta _{\mathbf{g},\mathbf{g^{\prime }}}+(\varepsilon
_s-\varepsilon _m)\frac{2\pi a^2}\sigma \frac{J_1(\mid \mathbf{g}-%
\mathbf{g^{\prime }}\mid a)}{\mid \mathbf{g}-\mathbf{%
g^{\prime }}\mid a}
\end{equation}
where \textit{a} is the cylinder radius at coordinate \textit{z} (Fig. 6), $J_1$ is the first-order Bessel
function, $\varepsilon _m$ is the dielectric constant of surrounding medium (here, vacuum), $\varepsilon _s$ is the
dielectric constant of PE, $\delta _{%
\mathbf{g},\mathbf{g^{\prime }}}$ is the Kronecker symbol and $%
\sigma $ is the primitive cell surface (Fig. 6). 
Due to the subwavelength size of the corrugation period, mode coupling (Eq. A5) gives rise to evanescent waves propagating along the surface.

Careful examination of Eq. A5 indicates that the coupling constant $\varepsilon _{\mathbf{g},\mathbf{g^{\prime }}}$ vanishes in two limit cases: (1) for low refractive index contrast,
\textit{i.e.} $\varepsilon _s\rightarrow
\varepsilon _m$, (2) in the topmost layers where
the cylinder radius $a$ becomes very small and ultimately tends to zero (note that lim$_{a\rightarrow 0}\frac{J_1(\mid 
\mathbf{g}-\mathbf{g^{\prime }}\mid a)}{\mid \mathbf{%
g}-\mathbf{g^{\prime }}\mid a}=\frac 12$). As a consequence, in both limit cases,
due to the extremely weak coupling, evanescent waves tend to vanish within the cavity. In the present study, the limit case 1 is always reached 
at wavelengths equal or shorter than 30 nm because of the dielectric properties
of PE, i.e. $\Re({\varepsilon_s})$ $\rightarrow$ $1$ and $\Im({\varepsilon_s})$ $\rightarrow$ $0$ (Fig. 7).
On the other hand, since evanescent waves propagate near the very top of the cones and given
the cones steepness thereof (i.e. $a$ $\rightarrow $ $0$)
the limit case 2 is always satisfied. These predictions need however to be verified by numerical simulations.

\begin{figure}
\centerline{\ \includegraphics[width=8cm]{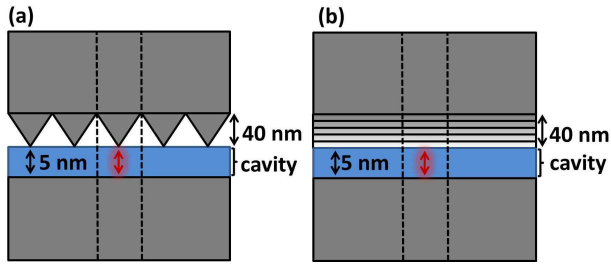}}
\caption{(Color online). Sketches of the configuration used in FDTD simulations for the cone array structure (a) and its corresponding EMA description (b). The red arrow denotes an oscillating dipole 
and dashed lines delimit the limits of the computational cell. One dipole is inserted in each cells and periodically repeated.}
\label{fig2}
\end{figure}

In order to probe the evanescent waves  inside the cavity, we 
numerically simulated, by finite difference time domain (FDTD) \cite{fdtd,fdtd2} home-made code, the diffracted field patterns which originate from oscillating dipoles inserted inside the cavity.
Such a model modelizes the electromagnetic field coming from quantum fluctuations inside the cavity.
Both
the actual 3D
structure (Fig. 8a) and its corresponding EMA description (Fig. 8b) were
simulated using cones height of 40 nm and a separation distance of 5 nm, as an illustration.
The electromagnetic responses of both configurations to dipole excitation were probed through normalized field intensity maps (Fig. 9). Maps are drawn for each computational cell (dashed lines on Fig. 8) and display only the diffracted field \textit{i.e.} dipole radiation removed.
\begin{figure}
\centerline{\ \includegraphics[width=8cm]{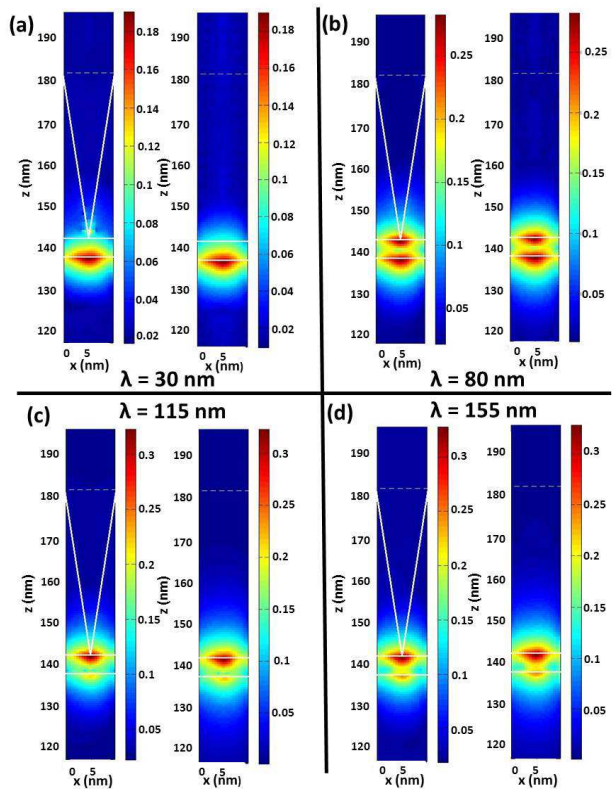}}
\caption{(Color online). Maps of the normalized intensity of diffracted field for different wavelengths in the three dimensional structure (left charts in each panel) and its EMA description (right charts in each panel).}
\label{fig2}
\end{figure}
The wavelengths of the radiating dipole (30 nm, 80 nm, 115 nm and 155 nm) were selected in order to sample the spectral range which is relevant to PE, according to its optical properties. At all these wavelengths except 30 nm, the limit case 1 is not reached
(Fig. 7) but the limit case 2 is reached near the top of the cones. In all cases, 
no significant differences are observed between the electromagnetic responses of the cone array
structure and its EMA description (Fig. 9,
relative error lower than 2\%).
This
result demonstrates that the fluctuations of the evanescent waves resulting from the coupling between
diffracted orders and specular order are extremely weak, which therefore plenty justifies the use of EMA in the present case. 

Note that, for a cavity made of materials with mobile charges, surface plasmon polaritons would produce strongly modulated evanescent waves which would dominate Lifshitz-van der Waals interactions \cite{plasmon, plasmon2}. In this case, the EMA description would obviously fail.
Since PE cannot endorse surface plasmon polaritons, the above mentioned problem is excluded here. 

In summary, the shallowness of evanescent waves fluctuations justifies the use of EMA in the present situation. The underlying physical reason is the weak modes coupling due to both the optical properties of PE and the steepness of the corrugation. The use of graded index profile for EMA description turns out to be reliable in rendering these shallow fluctuations. In addition, EMA avoids numerical stability issues while dealing with the direct computation of the scattering matrices of steep 3D structures made of materials with low-contrast optical indexes, while computing the Lifshitz-van der Waals force. This is a considerable advantage of the proposed method.

\section{Electrostatic potential energy calculation}

Hereafter, we evaluate the relative contribution of electrostatic forces to interfacial interaction in the case of PE slabs facing each other at very small separation distance (around 10 nm). Let us first consider two flat surfaces facing each other and separated by a
distance \textit{L} with a liquid as intervening medium. 
\begin{figure} [h!]
\centerline{\ \includegraphics[width=8 cm]{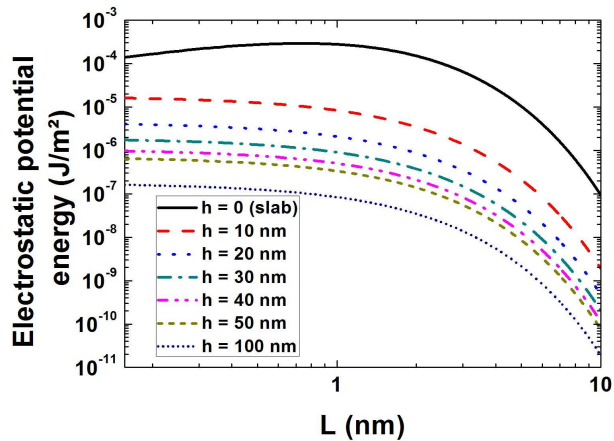}}
\caption{(Color online) Electrostatic potential energy between PE slabs as a function of cones height. Note the orders of magnitude on the Y-axis in comparison with Fig. 4.}
\label{fig1}
\end{figure}
The electrostatic potential
energy associated to the electrostatic double layer formed at both surfaces is given by \cite{1}:
\begin{equation}
W(L)=\varepsilon _r\varepsilon _0\kappa [2\psi _1\psi _2e^{-\kappa L}-(\psi
_1^2+\psi _2^2)e^{-2\kappa L}]
\end{equation}
where $\varepsilon _{r\text{ }}$is the static permittivity (dielectric constant) of the
liquid, $\varepsilon _0$ is the vacuum permittivity, $\kappa $ is
the inverse of the Debye length of the liquid and $\psi _i$ is the surface
potential of the i$^{th}$surface.

Replacing now one of the two surfaces by a
periodically nanostructured surface, here an array of cones, and using the Derjaguin
approximation \cite{1}, the electrostatic potential energy is given by:
\begin{eqnarray}
W(L)=2\pi \varepsilon _r\varepsilon _0\tan ^2(\alpha )
e^{-\kappa L}\\
\times [2\psi _1\psi _2- (1/4)(\psi _1^2+\psi _2^2)e^{-\kappa L}]/\kappa S \nonumber
\end{eqnarray}
\\
where $\alpha $ is the opening angle of the cone and $S=a_0^2\sqrt{3%
}/2$. 
It is noteworthy that the Derjaguin approximation is valid only if the
Debye length of the liquid is smaller than the lattice period \textit{a$_0$} of
the structured surface so that the electrostatic potential energy is not affected by coupling effects between cones. 

In the case study
presented in this article, the surface potential of both PE surfaces is equal to -
30mV \cite{1}, the static permittivity of water is equal to $78.2$ J m$^{-1}$V$^{-2}$ \cite{1} and the Debye length of water is equal to 1.5 nm \cite{1} (the Debye length of
groundwater is used for more realistic considerations). Note that
the value of the Debye length allows one to use Eq. B2 since the correlation length (\textit{i.e.} cones interdistance) is one order of magnitude bigger. The electrostatic potential energy for various cones height is shown
at Fig 10. The results of this calculation are commented in the article.


\begin{thebibliography}{99}
\bibitem{22bb} The results are presented in Fig. 4 for separation distances equal or shorter than the lattice parameter. Nevertheless, we checked that the same behavior of the interaction potential energy (i.e. transition from attractive to repulsive regimes) was obtained for separation distances larger than the lattice parameter.
\bibitem{22b} Having a perfectly flat interface is not experimentally achievable \cite{22a}. However, the experimentally observed deviations from the perfectly flat interface do not affect the concept presented here. Indeed, the Hamaker constant of a nanopatterned surface is modified at the global level only if the pattern exhibits short correlation lengths corrugations \cite{17}. Because the experimentally observed roughness occurs at long correlation length, the Hamaker constant of the ideally flat surface (in reality rough) is not affected.
\end{thebibliography}
\end{document}